# Regular Arrays of Pt Clusters on Alumina: a New Superstructure on $Al_2O_3$/$Ni_3Al$ (111)

Georges Sitja*,[§], Aude Bailly[†], Maurizio De Santis[†], Vasile Heresanu[§], Claude R. Henry*,[§]

[§] Aix-Marseille Université, CNRS, CINaM, F-13000 Marseille, France
[†] Univ. Grenoble Alpes, CNRS, Grenoble INP, Institut Néel, 38000 Grenoble, France

Corresponding authors:
Claude R. Henry: henry@cinam.univ-mrs.fr
Georges Sitja: sitja@cinam.univ-mrs.fr

Abstract:

Alumina ultrathin films obtained by high temperature oxidation of a $Ni_3Al$ (111) surface are a good template to grow regular arrays of metal clusters. Up to now two hexagonal organizations called 'dot' and 'network' structures have been observed with distances between clusters of 4.1 and 2.4 nm, respectively. In the present article we report on an investigation by in situ Grazing Incidence Small Angle X-ray Scattering (GISAXS), showing that Pt deposited at room temperature (RT) and for a low coverage forms a new hexagonal structure with a distance between clusters of 1.38 nm. For the first time, an assembly of tiny Pt clusters (1-6 atoms) with a very high density ($5.85 \times 10^{13}$ cm$^{-2}$) and presenting a good organization on an alumina surface, is obtained. This system could be used to investigate by surface science techniques the new emerging field of Single Atom Catalysis (SAC). By deposition at 573 K small Pt clusters are organized on the network structure. By deposition of Pt at 573 K on pre-formed Pd seeds, large Pt (Pd) clusters containing a hundred of atoms are organized on the dot structure and they remain organized up to 733 K. We show that the three structures are interrelated. The different organizations of the Pt clusters on the alumina surface are explained by the presence of 3 types of sites corresponding to different adsorption energy for Pt atoms.

## I. Introduction

Conventional supported model catalysts are obtained by growing metal clusters on metal-oxide surface[1-4]. However, nucleation of metal clusters occurs selectively on the randomly-distributed defects of the oxide surfaces, leading to a rather large size distribution (in the best case, 25% of the mean size) and the distance between the clusters is variable that can influence the local reaction kinetics[5]. These limitations can be overcome by using a substrate presenting a regular distribution of defects which play the role of a template to grow regular arrays of clusters with a sharp size distribution[5]. These arrays of clusters can be used, not only as model catalysts[5,6], but also for fundamental studies in nano-magnetism[7,8]. Three types of nanostructured surfaces have been recently used as templates: an ultrathin film of alumina obtained by oxidation of a $Ni_3Al$ (111) surface[9], a graphene monolayer (Gr)[10] or a monolayer of h-BN[11] grown on a metal single crystal. Regular arrays of Pt clusters have been obtained by deposition of Pt atoms on Gr/Ir (111)[12-16] or Gr/Rh (111)[17]. The distance between the clusters is 2-3 nm and the cluster size has been varied between 9 to 100 atoms. The cluster organization is stable up to 450K but CO adsorption at low pressure ($10^{-9}$ mbar) and RT induces coalescence[13,16]. Conversely, the clusters are stable at RT under $O_2$ or $H_2$ at $1 \times 10^{-8}$ mbar[14]. However under $1 \times 10^{-8}$ mbar of $O_2$ at 400 K, intercalation of oxygen atoms between the Gr and the metal substrate occurs, resulting in a disorganization of the cluster array and coalescence[18]. On h-BN/Ir (111) a good organization of Ir clusters has been obtained up to 700 K in ultra-high vacuum (UHV)[11]. To the best of our knowledge, only one STM image has been published for Pt clusters on h-BN/Rh (111), showing a broad size distribution of the clusters[19]. On $Al_2O_3$/$Ni_3Al$ (111) no results on regular array of Pt clusters have been reported, while arrays of V, Mn, Ag, Cu, Au[9], Pd[20], Fe[8,21] and Co[22] clusters have been obtained. However, only V, Pd and Cu atoms achieve a good organization. Arrays of bimetallic clusters have been first obtained for PdAu by successively depositing Pd and then Au[23,24]. In fact, Pd acts as seed for the subsequent deposition of Au. Seeding with Pd clusters has been used to obtain a good organization of Fe[8,21], Co[22] and Ni[25] clusters. The organization of the clusters occurs on specific nucleation sites forming two hexagonal superstructures which have been previously observed by STM on the clean alumina surface for different bias voltages. The first one, observed at V= 3.2 V, has a lattice parameter of 2.4 nm and is named 'network' structure while the second one, observed at 2.0 V, has a lattice parameter of 4.1 nm is named 'dot' structure[26]. The two superstructures are interrelated, the large mesh is ($\sqrt{3}$x$\sqrt{3}$) R30° relatively to the small one and the densities of sites are $6.5\times10^{12}$ cm$^{-2}$ and $1.95\times10^{13}$ cm$^{-2}$, respectively[9]. Pd clusters perfectly decorate the dot structure, while V clusters organize on the network structure[5]. The long range organization of the arrays has been studied *in situ* by GISAXS for Pd[6], Co (seeded by Pd)[22] and PdAu[27,28] clusters. The perfect organization of Pd and PdAu clusters is kept up to 600K in UHV[27] and in presence of CO and $O_2$ at a pressure of $1\times10^{-7}$ mbar[29]. The high stability of the cluster organization is certainly due to the particular surface structure of these films and probably also to their small thickness (0.5 nm). In summary, regular arrays of metal clusters on alumina/Ni$_3$Al (111) appear to be the best candidates to study accurately size effect in catalysis with surface science techniques, as it has been already shown for Pd and PdAu clusters[6,29-31]. It is rather astonishing that no attempt of Pt deposition on this alumina film has been reported yet, despite the great interest of Pt in heterogeneous catalysis. Here we present the first attempt to prepare regular arrays of Pt clusters on alumina ultrathin films on Ni$_3$Al (111) used as a template. The growth of the Pt clusters is followed *in situ* by GISAXS. This technique has the great advantage over STM that it reveals the organization of the clusters at the scale of the sample in a few minutes while several days would be necessary using STM. It is shown that depending of the growth conditions (temperature, Pt coverage) different organizations of the Pt clusters appear. In particular a new organization of Pt clusters containing few atoms and with a very high density is observed for the first time. We show how this new organization is related to the well-known network and dot structures of the alumina films on Ni$_3$Al (111).

## II. Experimental

The present experimental results have been obtained at the European Synchrotron Radiation Facility (ESRF, Grenoble, France) with the dedicated In Situ Nanostructures and Surfaces (INS) apparatus of the BM32 beamline. The surface of a Ni$_3$Al (111) single crystal (MaTeck, Jülich, Germany) was cleaned through successive cycles of ion bombardment and annealing at 1100 K. The sample was subsequently oxidized at an oxygen partial pressure of $5 \times 10^{-8}$ mbar at 1000 K during 20 minutes to obtain a nanostructured film of alumina[32]. Atomic deposition of palladium and platinum was performed by thermal evaporation from a water-cooled electron-beam evaporator. The palladium and platinum fluxes have been previously calibrated by using a quartz crystal microbalance. Sample preparation and metal deposition were conducted *in situ* under UHV in the same chamber used for the GISAXS measurements. The incident X-ray beam energy was set to 19.8 keV corresponding to a wavelength ($\lambda$) of 0.0626 nm. The incidence angle was equal to 0.16° (close to the angle of total external reflection for $Ni_3Al$ at this energy). GISAXS patterns were captured using a two-dimensional (2D) Eiger R1M camera from Dectris. The wavevector transfer **q** is defined by its three coordinates: $q_x$ and $q_y$, parallel with the sample surface, and $q_z$ perpendicular to it ($q_y$ and $q_z$ being both parallel with the detector plane). The in-plane ($2\Theta_f$) and out-of-plane ($\alpha_f$) scattering angles being small, the $q_y$ and $q_z$ components can be approximated by: $q_y \approx (2\pi/\lambda).\sin(2\Theta_f) \approx (4\pi/\lambda)\Theta_f$ and $q_z \approx (2\pi/\lambda).\alpha_f$. The surface unit cell of the dot structure is hexagonal with the norm of its basis vector equal to 4.1 nm, leading to a distance between nanoparticles rows of D = 4.1×$\sqrt{3}$/2 = 3.55 nm. The regular organization of the clusters on the alumina film gives rise to sharp diffraction rods on the GISAXS patterns, perpendicularly to the sample surface. For any of the superstructures several diffraction lines corresponding to the nodes (1 0), (2 0), (3 0)... but also (1 1), (2 2), (3 3)... are expected to appear, as long as the sample azimuth has been correctly set. In some cases, the 2D camera has also been shifted with respect to the direct and specular beams in order to record the relevant diffraction peaks. The closest scattering rod from the (0 0) specular reflection corresponds to the node (1 0) and is therefore called "1st order" or "order 1". (2 0) and (3 0) nodes lead to rods that are 2 or and 3 times farther than order 1, and we will label these lines as order 2 (or 2nd order) and order 3 (or 3rd order). The nodes (1 1) and (2 2) of the reciprocal lattice lead to rods respectively $\sqrt{3}$ and $2\sqrt{3}$ farther than order 1. These lines are called order $\sqrt{3}$ and order $2\sqrt{3}$.

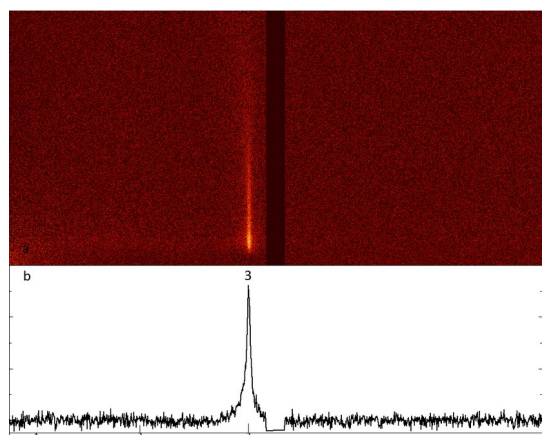

*Figure 1: 0.1 ML Pt deposited at RT (a) GISAXS pattern at ω ≈ 0° (b) intensity profile. (The exact value of ω for this figure is -1.5° in order to maximize the intensity of the third order peak. For this value if the first and second order peaks would exist they would be visible as seen in Supplementary Information.)*

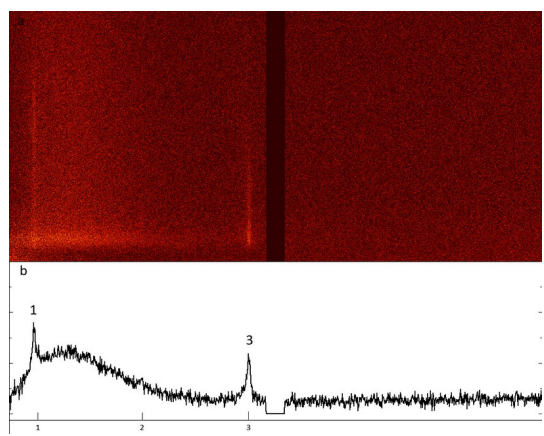

*Figure 2: 0.35 ML Pt deposited at RT (a) GISAXS pattern at ω ≈ 0° (b) Intensity profile. (The exact value of ω is -1.5°)*

To record enough signal in the relevant diffraction rods, the azimuth angle ($\omega$) is slightly tilted from the natural directions of the clusters array on the surface to form the corresponding Bragg angle. However, the azimuth to record order 1, 2 and 3 is nearly 30° away from the one to record $\sqrt{3}$ and $2\sqrt{3}$ rods. Hereafter, we will simplify using only 0° and 30° to label the GISAXS data (the exact value of $\omega$ is indicated in the caption for each figure). 0° will stand for order 1, 2, 3... and 30° for $\sqrt{3}$, $2\sqrt{3}$...

The values of $\omega$ which maximize the intensity of the different peaks and the width of the peak intensity as a function of $\omega$ are calculated in Supplementary Information. The experimental variation of the order 1 peak as a function of $\omega$ is displayed in Supplementary Information (Fig. SI1). In the intensity profiles the peak intensity is integrated within a band containing 220 lines from the 1030 lines of the experimental image.

## III. Results

### 3.1 RT deposition

Figure 1a shows the GISAXS pattern ($\omega$ = 0°) after deposition of 0.1ML Pt at RT on the alumina film. Only one peak is visible corresponding to the 3rd order in reference to the pattern obtained with the dot structure. On the intensity profile corrected from the baseline of the clean alumina (Figure 1b), the background is flat indicating that all the scattered intensity from the Pt clusters is in the third order peak. No disorder nor coalescence are observed. As the first and second order peaks are not present, the observed peak corresponds to a unit cell 3 times smaller than the unit cell of the dot structure (in the direction $\omega$ = 0°). At $\omega$ = 30°, no peak is observed on the GISAXS pattern (See Supplementary Information, fig. SI2), but a very small peak is visible on the intensity profile at the position $3\sqrt{3}$ (see figure 5 for the indexation of the peaks). As the $\sqrt{3}$ and $2\sqrt{3}$ peaks are not visible, we deduce that the Pt clusters do not sit on the network structure. The $3\sqrt{3}$ peak corresponds to a second order scattering peak of the new structure. From these observations we conclude (see discussion) that the Pt clusters are on a hexagonal structure parallel to the dot structure, but with unit cell vectors 3 times smaller (1.37 nm) than those of the dot structure. We call this new structure "dot/3". The density of clusters is therefore 9 times larger than for the dot structure and equal to 5.85x10$^{13}$ cm$^{-2}$. For 0.1 ML of Pt, the clusters contain 2.6 atoms on average. Similar GISAXS patterns have been clearly observed for 0.05 (1.3 atoms) and 0.15 ML (3.9 atoms)

Figure 2 displays the GISAXS pattern for 0.35 ML of Pt deposited at RT. We now clearly see the 1st and 3rd order peaks, while the 2nd order peak is hardly visible on top of a broad diffuse component, characteristic of the disorder due to the coalescence of the clusters. No sharp peak is seen at ω = 30°, but only a broad feature in the background. From simulation of GISAXS for a pure dot structure we expect 1st, 2nd and 3rd peaks at ω = 0° and √3 peak at ω = 30°. For a pure network structure we expect 3rd order peak at ω = 0° together with √3 and 2√3 peaks at ω = 30°. Thus we have not a pure structure but we can explain the experimental observations by the following remarks. The presence of the peak 1 shows that locally some clusters are on the dot structure and the fact that the peak 3 is much intense than the peak 2 (taking into account that the Debye-Waller factor is proportional to Q²) proves that a large fraction of the clusters are still on the "dot/3" structure at 1.37 nm. The absence of √3 peak at ω=30° shows that locally the clusters are not organized on the network structure. The presence of a wide and intense diffuse peak indicates that the collection of clusters is no longer well-ordered.

### 3.2 Pt deposition at 573 K

On figure 3, the first row displays the GISAXS patterns in two directions (ω = 0° and ω = 30°) for 0.08 ML Pt deposited at 573 K. At ω = 0°, no clear peak is visible, while at ω = 30° a √3 peak is present. These observations indicate that only the network structure is occupied by the Pt clusters which contain 6.3 atoms on average. The density of clusters is $1.95 \times 10^{13}$ cm$^{-2}$. It should be noticed that in the configuration of the camera used for this series of experiments, the peaks 3 and 2√3 are out of the field of view of the camera.

After a second deposition of 0.08 ML (total 0.16ML Pt) the √3 peak becomes more intense (figure 3, second row) and a very small 1st order peak can be detected close to the baseline. We conclude that for 0.16 ML Pt, the clusters (≈12.6 atoms per cluster) still sit on the network structure, but there is a slight imbalance (in size) between clusters that are on the sites common with the pure dot structure (site A) and those that are specific of the network structure (site B). Coming back to the profile for the initial 0.08 ML deposit at ω = 0°, we can possibly see a tiny peak at the position of the order 1 peak, which is almost merged in the background. This tiny peak would indicate that at 0.08 ML, there is already a small dissymmetry between the size of the clusters in sites A and B, but in both cases the Pt clusters are organized on the network structure.

After deposition of 0.85 ML of Pt at 573K the imbalance between the two kinds of clusters is still noticeable but also a wide diffuse peak due to disorder and coalescence is observed (figure not shown).

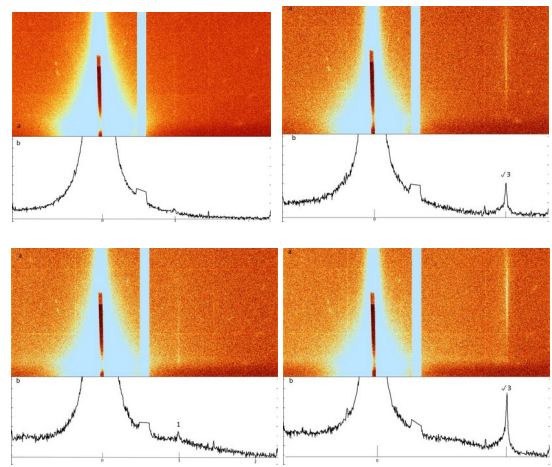

*Figure 3: Pt deposition at 573K. First row: 0.08 ML, ω ≈ 0° (left) and ω ≈ 30° (right). Second row: 0.16 ML, ω ≈ 0° (left) and ω ≈ 30° (right). (a) and (b) represent the GISAXS pattern and the intensity profile, respectively. (The exact value ω are -0.5°, 28.8° for the left and right parts of the figure, respectively)*

### 3.3 Pd seeding

In the previous section we have seen that it is not possible to organize Pt clusters containing about a hundred atoms on the network structure. In order to obtain a regular array of large Pt clusters we tried to use Pd seeding on the dot structure like for Au, Fe, Co and Ni[21-23,25].

Figure 4 (first row) shows GISAXS patterns for 0.1 ML of Pd deposited at 363 K. At ω = 0°, 1st and 2nd order peaks are clearly seen proving that the clusters containing 23.5 Pd atoms on average are organized on the dot structure. After heating the sample at 573 K, the clusters stay organized on the dot structure (right image on the first row). Then, 0.48 ML of Pt was deposited at 573 K on the pre-formed Pd clusters. GISAXS pattern (figure 4, left of second row) shows that the clusters containing 23.5 Pd atoms and 113 Pt atoms stay organized on the dot structure. The large augmentation of the intensity of the first order peak is due to the important increase of the cluster volume (almost a factor 6). The vanishing of the second order peak is also due to the large increase of the cluster size, which decreases the visibility of the high order peaks through the modification of the cluster form factor. After deposition of 0.78 ML of Pt on the Pd seed clusters (23.5 Pd atoms and 184 Pt atoms), the metal clusters stay organized on the dot structure although some disorder starts to appear (figure 4, right part of second row). After annealing the sample successively at 711K and 733K, GISAXS patterns show that the organization on the dot structure is kept up to 733K. Nevertheless, the intensity of the GISAXS first order peak decreases when temperature increases, indicating that the organization becomes less perfect.

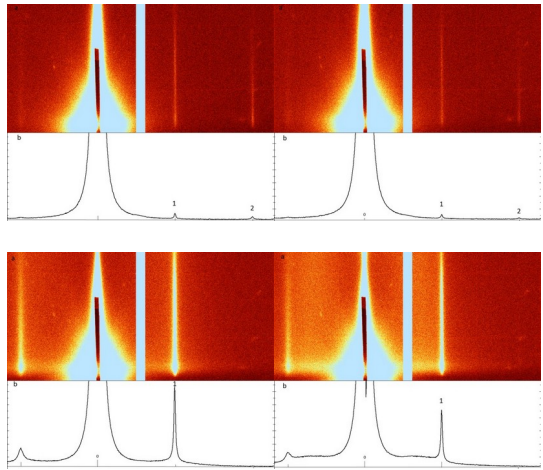

*Figure 4: GISAXS pattern and corresponding intensity profiles for Pt cluster arrays seeded with Pd. First row, left: 0.1 ML Pd deposited at 363K and right: after annealing at 573 K. Second row, left: after deposition of 0.48 ML Pt at 573K on the Pd seeds and right: after deposition of 0.78 ML Pt at 573K on the Pd seed. (a) and (b) represent the GISAXS patterns at $\omega \approx 0°$ and the corresponding intensity profile, respectively. (The exact value of $\omega$ is -0.5°)*

## IV. Discussion

**4.1 Origin of the new structure**

The first study of ultrathin alumina films on a $Ni_3Al$ surface was made by the group of U. Bardi[33,34]. The films were prepared by oxidation at 900 K of a polycrystalline $Ni_3Al$ surface. By LEED they observed a surface mesh at ≈3Å attributed to oxygen atoms of the alumina film with preferential (111) orientation. The thickness of the films has been determined by XPS to be 0.5 nm[33,34]. Later, C. Becker et al.[35] prepared the alumina film on $Ni_3Al$(111) by oxidation at 300K and annealing at 1000K. They provided the first STM images of the alumina film[35]. They observed two superstructures at 2.5 nm and 4.6 nm which appear at different bias voltages respectively 3.1 and 2.2 V[35]. A few months later, J.A. Kelber and collaborators, used the same preparation method and observed by STM at RT a hexagonal lattice at 3 Å they identified as O surface atoms[36]. By direct oxidation at 1000K and STM observation at low temperature, more precise images of the alumina films were obtained[26]. At 3.2 V a hexagonal arrangement of "dark holes surrounded by smaller hexagonal rings of bright dots"[26] was observed characterizing the network structure. At 2.0 V a hexagonal arrangement of bright dots was observed which is characteristic of the dot structure[26]. The unit cell vectors were more precisely determined as 2.35 nm for the network structure and 4.16 nm for the dot structure[26]. At this time the question was: do these structures have a pure electronic origin or correspond to topographic surface structures? Non-contact Atomic Force Microscopy (nc-AFM) is the best suited technique to answer this question. In 2006 some of us observed the surface of the alumina films by nc-AFM[32]. Atomic resolution showed the hexagonal surface lattice at 2.9 Å attributed also to O atoms. The self-correlation of the atomic resolution image showed modulation, revealing the network and the dot structures with parameters of 2.39 nm and 4.1 nm, respectively[32]. The modulation was explained by distortion of the atomic lattice due to the non-perfect accommodation with the $Ni_3Al$ substrate[32]. nc-AFM images in the damping mode clearly showed dot and network superstructures which are defined by two types of features[32]. Hexagons formed by 6 spots separated by 8 Å appear around the nodes of the dot structure, and the unit mesh of the network structure is formed by 3 single spots and one hexagon in common with a node of the dot structure unit mesh[32]. From the Fourier transform of the damping AFM image obtained on a single domain it was deduced that the three different features form three commensurate lattices[32]. We reproduce this Fourier transform (not shown in reference 32) on figure 5. In this figure one recognizes the reciprocal unit-mesh of the dot structure (a = 4.14nm), the one of the network structure ($a/\sqrt{3}$ =2.39 nm) and the one of the smallest structure forming the 6 hexagonal spots ($a/3\sqrt{3}$ = 0.797nm) that we call "network/3". We can also recognize the unit cell corresponding to the new structure observed for Pt deposition at RT and at low coverage with a unit cell parameter equal to a/3 = 1.38 nm. As said we name this new structure: dot/3. All the four structures are interrelated. The characteristic distance of this new structure (dot/3) corresponds in fact exactly to the one observed in STM images at V = 3.1/3.2 V (in fig. 3a of Reference 26 and on fig. 3b or 3d of Reference 21) between the 6 bright spots forming hexagons.

| Superstructure | dot | network | dot/3 | network/3 |
|---|---|---|---|---|
| Lattice parameter (nm) | 4.14 | 2.39 | 1.38 | 0.797 |
| Cluster density (cm$^{-2}$) | 6.5x10$^{12}$ | 1.95x10$^{13}$ | 5.85x10$^{13}$ | |
| Relationship | 1x1 | (1/√3x1/√3)R30° | 1/3x1/3 | (1/3√3x1/3√3)R30° |

Table 1: Characteristics of the hexagonal superstructures observed on the alumina films

In another nc-AFM study[37] the atomic hexagonal lattice forming the top surface of the films was also seen with a parameter of 2.9 Å. The network and the dot structures were also recognized. By inverse Fourier transform from selected spots, a hexagonal lattice at 5.1 Å was observed and associated with the Ni$_3$Al substrate lattice[37]. The characteristic of the different superstructures are presented in Table 1. It is interesting to notice that from the reciprocal lattices displayed in fig. 5, one can easily predict the GISAXS peaks that must be observed for the different structures in the two azimuths (ω=0° and ω=30°). We have simulated the GISAXS patterns for the different structures and also considered different occupations of the A, B and C sites (see next paragraph). A good agreement between the simulations and the experimental GISAXS patterns for 0.1 ML of Pt deposited at RT was obtained only for a full occupation of the dot/3 structure (see Supplementary Information, 3$^{rd}$ section).

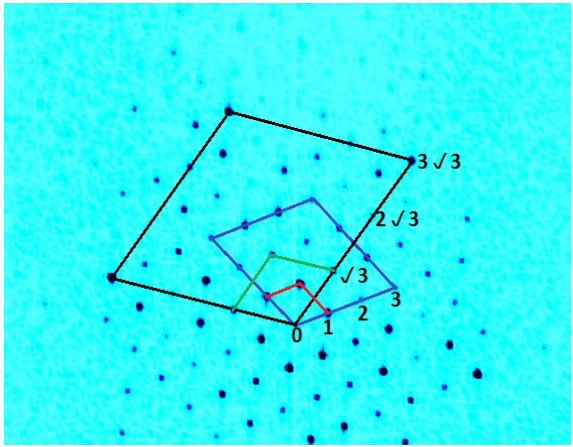

Figure 5: Fourier transform of the nc-AFM image (Fig. 1 of Reference 32) in damping mode of a single domain of the Al2O3/Ni3Al (111) films (the contrast has been inverted and enhanced). The red, green, blue and black unit meshes correspond to the dot, network, dot/3 and network/3 structures, respectively.

**4.2 Nature of the different sites on the surface of the alumina films**

In our previous nc-AFM study of the alumina surface we defined two types of sites (A and B)[32] (see also Supplementary Information). The A sites correspond to the nodes of the dot structure, while the unit mesh of the network structure is composed by one A site and three B sites. Now we define two additional sites (C and D). The C sites are at 1/3 and at 2/3 of the lattice vectors of the unit mesh of the dot structure. Therefore, the nodes of the unit mesh of the "dot/3" structure are one A site and three C sites (see figure 6). Finally D sites correspond to the six dots around an A site observed by nc-AFM[32]. The nodes of the unit mesh of the network/3 structure are composed of one A site and three D sites. The different superstructures unit cells, together with the different types of sites, are schematically represented on figure 6. A real space representation of the different sites can be obtained from the previous nc-AFM[32] and is given in figure SI5 in the Supplementary Information. It is interesting to notice that the B, C and D sites form hexagons of decreasing size around the A sites and successively rotated by 30°.

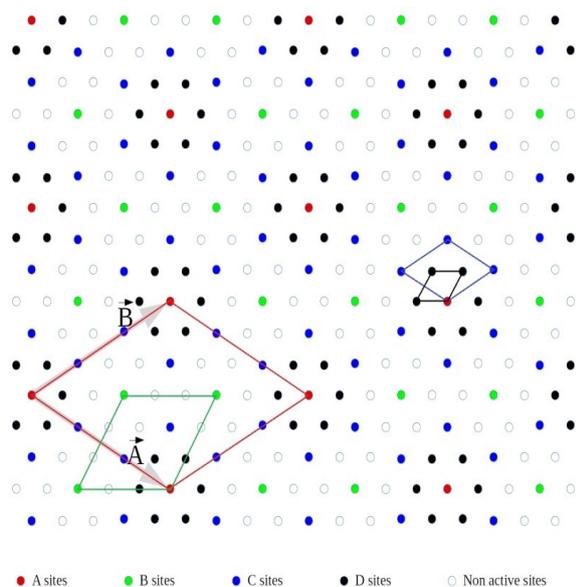

- A sites
- B sites
- C sites
- D sites
- Non active sites

*Figure 6: The unit mesh of the dot, network, dot/3 and network/3 superstructures are represented in red, green, blue and black lines, respectively. The A and B vectors (in grey) are the unit mesh vectors of the dot structure which is the only true crystallographic surface structure. The A, B, C, D sites are the red, green, blue and black dots, respectively.*

An atomic model based on the comparison between near atomic resolution STM images and Density Functional Theory (DFT) calculations has been given by M. Schmid *et al*.[21]. Following this model the nodes of the dot structure (A sites) correspond to holes through the alumina films which are strong adsorption sites for metal atoms like Pd. Further calculations with this model structure showed that it is also the case for Co, Fe, Ni, Cu, Ag and Au atoms[38]. The high adsorption energy of metal atoms on the holes is mainly due to the fact that the alumina film does not totally screen the influence of the metallic $Ni_3Al$ substrate. No calculation was made for Pt, but we can expect that it is also the case. Looking carefully at the atomistic model (Fig. 2 of Reference 21) the B sites forming the network structure correspond to the center of a triangle formed by 3 surface oxygen atoms. The C sites are also found approximately at the center of a triangle formed by 3 surface oxygen atoms, but the environment of these 3 oxygen atom is really different for the two types of sites (B and C). Therefore, we expect that the adsorption energy of a Pt atom is different on these two sites. Thus, we can expect that the adsorption energy on C sites would be close to those on B sites. But to explain the fact that at RT the Pt atoms populates the new structure (dot/3) and that at 573 K only the network structure is occupied we have to assume that the adsorption energy on C sites is smaller than on B sites. The sites corresponding to the smallest structure (D sites) appear to be close to a site between 4 oxygen atoms forming a square. The adsorption energy of a Pt atom on D sites must be significantly weaker than on the other sites, because at RT they are not occupied by Pt atoms. We assume that there is a hierarchy in the adsorption energy of Pt atoms on the different sites; this value decreasing in the order A>B>C>D. From this simple argument we can tentatively explain the experimental observations. At RT, the A, B and C sites are occupied by small Pt clusters (typically less than 6 atoms). From a cluster size of 6 atoms, the adsorption energy of the cluster is not large enough to stabilize the cluster on C sites and they start to diffuse and coalesce leading to some disorder. At 573K, the C sites cannot stabilize Pt atoms. Then only B and A sites are occupied and form the network structure for Pt clusters containing up to ≈ 13 atoms. At larger sizes Pt clusters are no longer stable on B sites leading to a mixture of dot and network structures. Some disorder is also present and increases with the Pt coverage. In order to have a regular array of larger Pt clusters we have to use Pd seeding to form Pd-seeded Pt clusters on the dot structure. In that case, Pd is deposited at 363 K and Pt is subsequently deposited at 573 K. Even if small Pt clusters are originally formed on B sites, they diffuse and coalesce with Pt/Pd clusters on A sites, when they reach a critical size of about 13 atoms and therefore become stable. The Pd-seeded Pt clusters are stable on A sites up to 730 K.

### 4.3 Comparison between Pd and Pt

For Pd deposition at RT, the dot/3 structure was never observed, contrary to Pt. This difference suggests that the adsorption energy of Pd atoms on the different sites is weaker than for Pt. Several DFT calculations of the adsorption of Pt and Pd atoms on clean α-alumina (0001) surfaces have been published[39,40]. These calculations show that both Pt and Pd bind to oxygen atoms of the alumina surface, but more strongly for Pt (1.99 eV/1.82 eV) than for Pd (1.47 eV/1.19 eV). *Ab initio* calculation of the adhesion energy of Pt and Pd reaches the same conclusion[41]. On -alumina, Pt and Pd atoms are found to bind preferably to oxygen with a stronger binding energy in the case of Pt[42,43]. It is also found that the diffusion energy of metal atoms on -alumina is very weak (0.3-0.5 eV)[42,43]. From these calculations, we can understand that, at RT, Pt atoms can be stable on C sites and not Pd atoms. Moreover, the weak calculated diffusion energy shows that already at RT Pd and Pt atoms diffuse rapidly on the surface, as soon as they escape from the B and C sites. For a very thin oxide film supported on a metal substrate, like the alumina used in this study, one would expect an increase of the adsorption energy of a metal atom in comparison with bulk alumina[44]. This is true for Pd on A site but an adsorption energy similar to bulk alumina is calculated[38] for Pd on B sites (1.18 eV).

## V. Conclusion

For the first time, regular arrays of Pt clusters have been obtained by deposition of Pt atoms on an ultrathin alumina film grown on $Ni_3Al(111)$. Depending on the deposition temperature different structures are obtained. At room temperature a new structure that we call "dot/3" is observed. Up to now, this new organization had never been observed for the growth of any other metal clusters on this alumina film. The density is 9 times larger than the density for the dot structure that gives $5.85 \times 10^{13}$ clusters cm$^{-2}$. Only very small Pt clusters (less than 6 atoms) can be organized on this structure. This system is particularly well suited to study 'single atom catalysis'[45] by surface science techniques like molecular beam techniques[29]. Indeed from single atom catalysts supported on an oxide powder, it is rather difficult to accurately know the geometry of the site where single atoms of catalytic metal are anchored. The use of a flat single crystal under UHV allows to use surface science techniques like LEED, STM, nc-AFM… to determine the geometry of adsorption sites. For such studies, a few systems have been recently used, like single Au, Ag, Ni and Pt atoms on $Fe_3O_4$ (001)[46-49] or Pt on CuO/Cu(110)[50]. The new system investigated here has the advantage of presenting a regular organization of few atoms Pt clusters on an alumina surface, which is the most common substrate used in SAC.

By deposition at 573 K, a regular array of Pt clusters, containing 13 atoms and forming the network structure ($1.95 \times 10^{13}$ clusters.cm$^{-2}$), is observed. Larger clusters containing 50 atoms are no longer ordered on the network structure. In order to obtain a regular array of large Pt clusters it is necessary to seed the surface with small Pd clusters. By deposition of Pd at 363 K a regular array of Pd clusters is formed on the dot structure ($6.5 \times 10^{12}$ clusters.cm$^{-2}$). The subsequent deposition of Pt atoms at 573 K leads exclusively to the growth of the Pd seeded clusters. The organization of Pd seeded-Pt clusters on the dot structure is stable up to 730 K.

On the alumina film four surface structures are observed: the dot structure, the network structure, the dot/3 structure and the network/3 structure. These 4 structures are deeply interrelated. Depending on the growth conditions, the organization of Pt clusters on the three larger structures is observed. The formation of these different organizations is tentatively explained by the presence of 3 types of sites: A, B and C having respectively a decreasing interaction with Pt atoms. The dot structure is formed exclusively by A sites, the network structure is formed by A and B sites and the dot/3 structure is formed by A, B and C sites. The validity of this model will be tested by further experiments in STM and simulations of the GISAXS patterns.

## Supporting Information

1. Optimal values of ω for the different GISAXS peaks

2. GISAXS pattern, in the direction ω = 0°, for 0.1 ML of Pt deposited at RT

3. Simulation of GISAXS pattern

4. Definition of the different adsorption sites on the alumina film/$Ni_3Al(111)$.

## ACKNOWLEDGMENTS

Financial support through ANR EQUIPEX ANR-11-EQPX-0010 and beam time on the French CRG-IF beamline at the ESRF are acknowledged. The authors thank the beamline staff.

## Supplementary Information

1. Optimal values of ω for the different GISAXS peaks.

The value ω = 0° corresponds to the symmetrical position where we see the peaks of orders 1 and -1 with the same intensity. In the conditions of the experiment (λ = 0.0626 nm) the calculated omega values to have the maximum intensity for a given peak (Bragg condition) are: -0.5°, -1°, -1.5° for 1,2, 3 orders and 29.13°, 28.27°, 27.4° for √(3), 2√(3), 3√(3).

Relation used for the calculation of the omega range for which a spot is observed, with respect with omega rotation

$$\omega = acos\left(\frac{q^2 - 2*\Re*Rd - Rd^2}{2*\Re*Rd}\right) - acos\left(\frac{q^2 + 2*\Re*Rd - Rd^2}{2*\Re*Rd}\right)$$

Where :

q = 2*π/d(hkl)
Re = radius of the Ewald's sphere = 2*π/λ
Rd = radius of the diffraction spot

Numerical application, for the experimental condition used

| order | q (nm$^{-1}$) | Re (nm$^{-1}$) | Rd (nm$^{-1}$) | ω (deg) |
|---|---|---|---|---|
| 1 | 1.75 | 100.37 | 0.04 | 2.4963 |
| 2 | 3.50 | 100.37 | 0.04 | 1.2482 |
| 3 | 5.26 | 100.37 | 0.04 | 0.8323 |
| √(3) | 3.04 | 100.37 | 0.04 | 1.4412 |
| 2 √(3) | 6.07 | 100.37 | 0.04 | 0.7209 |
| 3 √(3) | 9.11 | 100.37 | 0.04 | 0.4808 |

Figure SI1 shows an experimental intensity profile for the first order peak measured at the Yoneda position for a Pd deposit. The experimental FWHM (2.5°) is in perfect agreement with the calculated one.

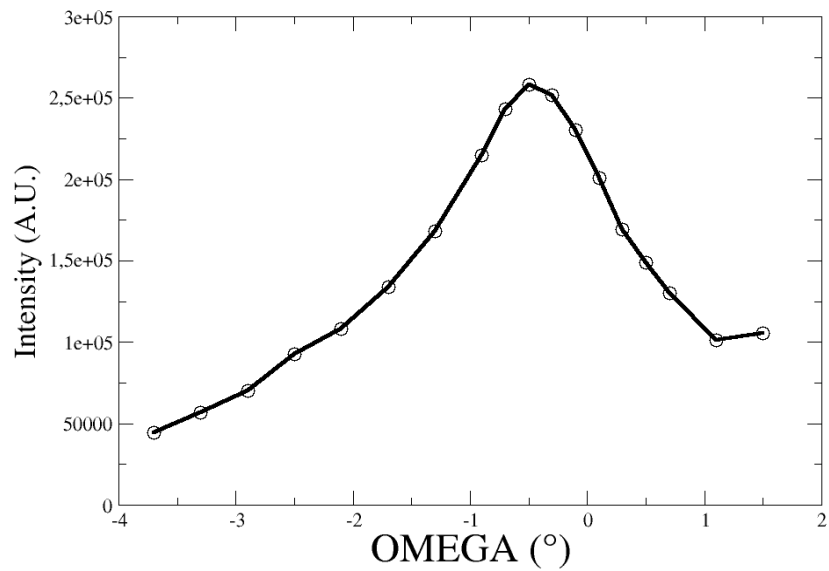

Figure SI1: Intensity of first order peak versus ω for a Pd deposit.

2. GISAXS pattern for 0.1ML Pt deposited at RT obtained at ω = 29.13°.

The peak of order $\sqrt{3}$ and $2\sqrt{3}$ are not visible. A tiny peak of order $3\sqrt{3}$ (indicated by an arrow) is visible on the smoothed intensity profile (green line).

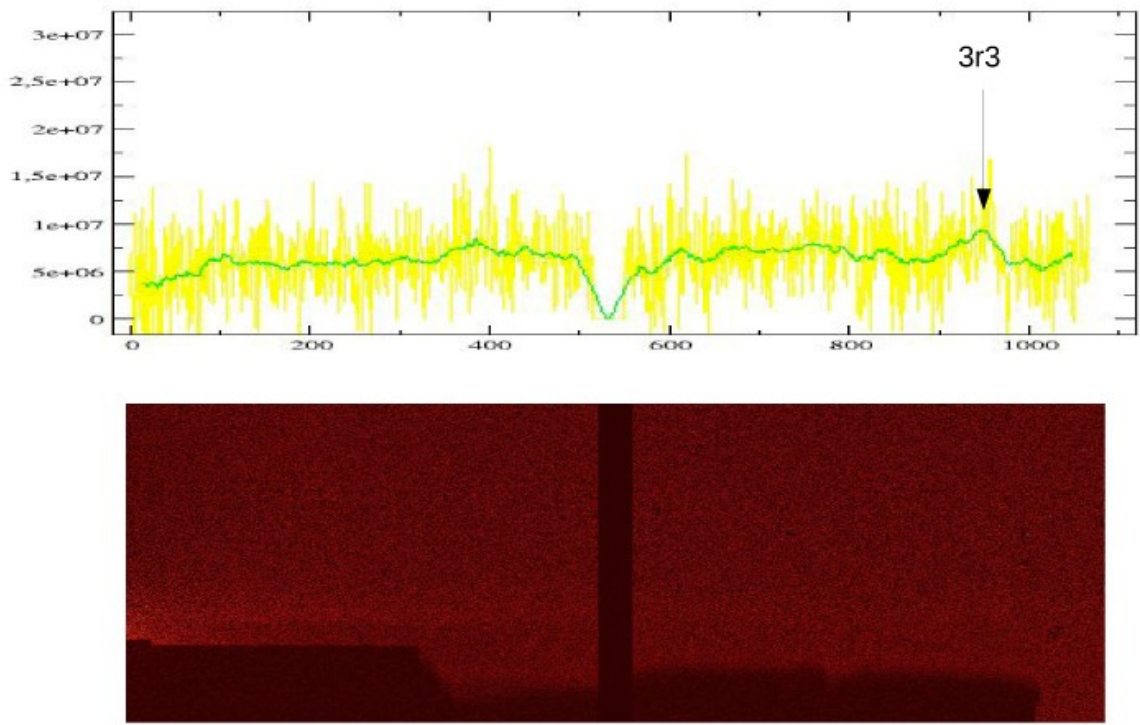

Figure SI2: Intensity profile (top) and GISAXS pattern (bottom).

3. Simulation of the GISAXS pattern

Figure SI3 shows a simulation of the intensity at the Yoneda peak for GISAXS peaks of order 1, 2 and 3 (black, red and blue curve, respectively) as a function of ω. The simulation is made for a distribution of hemispherical clusters containing 22 atoms (diameter of 1.1 nm, equivalent to a Pt deposit of 0.1 ML) which are positioned on A sites (dot structure) with an accuracy of 0.1 nm. If we chose the value ω = -1.5° to have the maximum intensity for order 3 (condition of figure 1a in the main text) we see that the peaks of order 1 and 2 must be clearly visible. The same simulation has been made for clusters organized on the network structure and they show that the peak of order $\sqrt{3}$ must be visible. In the experiments for 0.1 ML Pt deposited at RT only the peak of order 3 is visible while the peaks of order 1,2 and $\sqrt{3}$ are not visible then it can be safely concluded that the Pt clusters are not on the dot nor on the network structure.

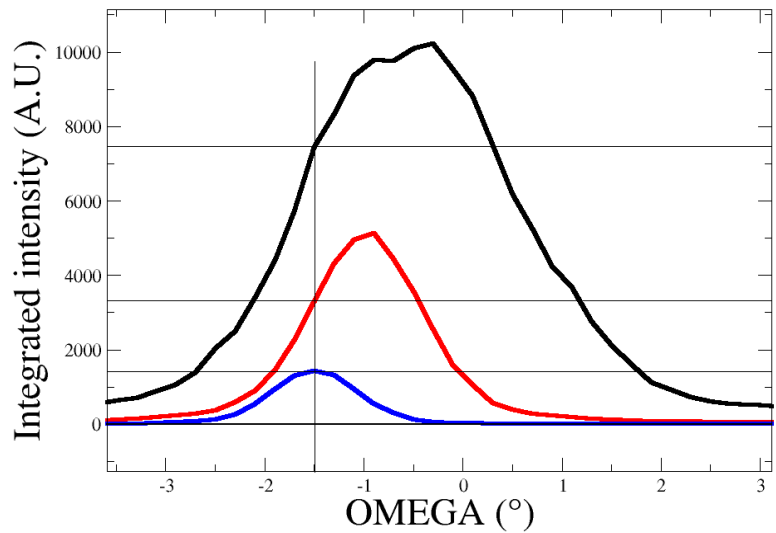

Figure SI3: Intensity of peaks of order 1, 2 and 3 as a function of ω simulated for Pt clusters containing 22 atoms organized on the dot structure.

Only the simulation of Pt clusters on A, B and C site with a full occupancy corresponding to the dot/3 structure (see Figure SI4) gives GISAXS patterns in agreement with experiment for ω= -1.5° and ω = 29.13° i.e. absence of peak 1 and 2 in one azimuth and absence of $\sqrt{3}$ and $2\sqrt{3}$ in the other one: only the peak of order 3 is visible.

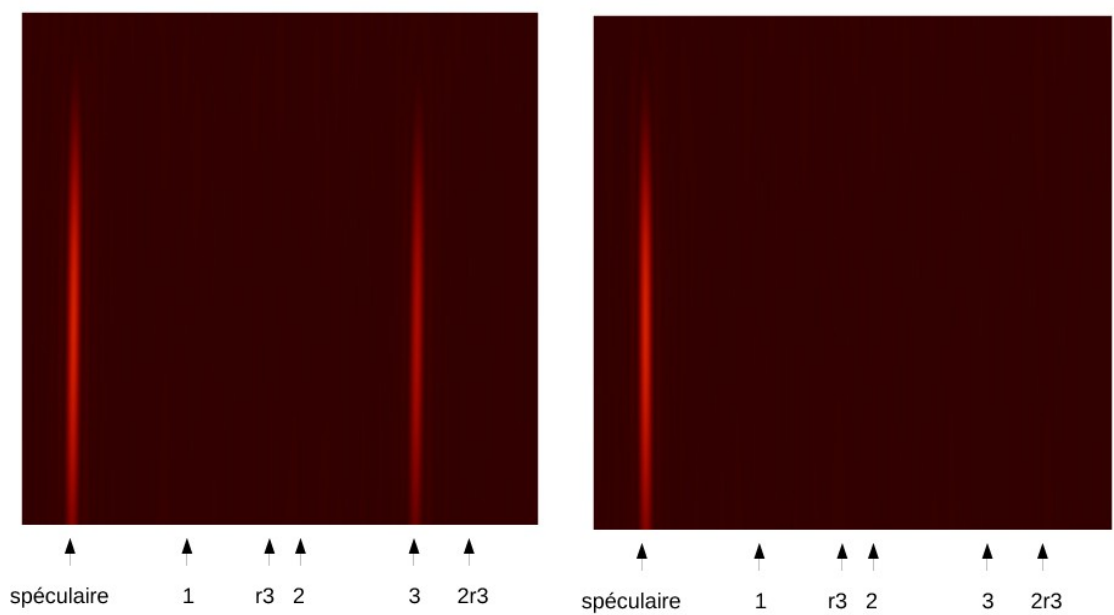

Figure SI4: GISAXS Simulation of 0.1 ML of Pt on the dot/3 structure. On the left w= -1.5 and on the right w = 29.13°

4. Definition of the different adsorption sites on the alumina film on Ni$_3$Al (111).

In a previous nc-AFM study [1], damping images evidenced two features: 6 dark dots forming an hexagon centered on the nodes of the dot structure (defined as A sites) and single dark dots (defined as B sites) forming together with one A site the unit mesh of the network structure. The power spectrum (obtained by Fourier transformation) of this nc-AFM image gives the reciprocal lattice of the superstructure of the alumina film (see figure 5 of the main text). By selecting the peaks from the power spectrum and applying a back Fourier transformation, one obtains figure S1 where the lattice formed by the six black dots around the A sites is propagated on the whole image. Now it is easy to recognize the new dot/3 superstructure, corresponding to near RT deposition of Pt at low coverage, and the network/3 superstructure. Then we can easily identify the four types of sites: A, B,C and D.

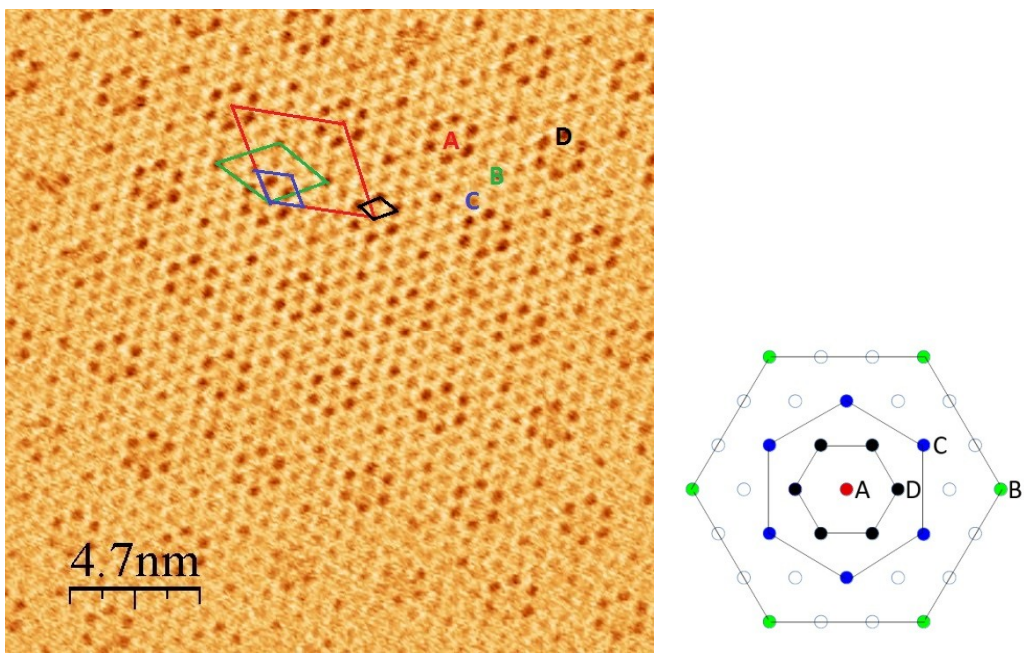

Figure SI5: Inverse Fourier transformation of the power spectrum presented in figure 5 obtained by selection of the peaks corresponding to the four structures discussed in section IV. The unit meshes of the dot, network, dot/3, and network/3 structures are represented in red, green, blue and black respectively. The A, B, C and D sites are indicated by the center of letters colored in red, green, blue and black, respectively. On the right we see that the different sites (A red, B green, C blue and D black) form hexagons centered on A sites.

[1] Hamm, G.; Barth, C.; Becker, C.; Wandelt, K.; Henry, C.R.; Surface structure of an ultrathin alumina film on Ni$_3$Al(111): A dynamic scanning force microscopy study. *Phys. Rev. Let.* **2006**, 97, 126106.